\documentclass[preprint,prl,aps,showpacs,amsmath,amssymb]{revtex4-1}

\usepackage{graphicx}
\usepackage{epstopdf}
\usepackage{braket}
\usepackage{float}
\usepackage[euler]{textgreek}
\begin{document}

\title{Quasiparticle Interference in ZrSiS: Strongly Band-Selective Scattering Depending on Impurity Lattice Site}

\author{Christopher J. Butler$^{1,}$\footnote{These authors contributed equally to this work.}$^{,}$\footnote{cjbutler@ntu.edu.tw}}
\author{Yu-Mi Wu$^{1,}$\footnotemark[1]}
\author{Cheng-Rong Hsing$^{2}$}
\author{Yi Tseng$^{1}$}
\author{Raman Sankar$^{3,4}$}
\author{Ching-Ming Wei$^{2,}$\footnote{cmw@phys.sinica.edu.tw}}
\author{Fang-Cheng Chou$^{4,5,6}$}
\author{Minn-Tsong Lin$^{1,2,7,}$\footnote{mtlin@phys.ntu.edu.tw}}

\affiliation{$^{1}$Department of Physics, National Taiwan University, Taipei 10617, Taiwan}
\affiliation{$^{2}$Institute of Atomic and Molecular Sciences, Academia Sinica, Taipei 10617, Taiwan}
\affiliation{$^{3}$Institute of Physics, Academia Sinica, Taipei 11529, Taiwan}
\affiliation{$^{4}$Center for Condensed Matter Sciences, National Taiwan University, Taipei 10617, Taiwan}
\affiliation{$^{5}$National Synchrotron Radiation Research Center, Hsinchu 30076, Taiwan}
\affiliation{$^{6}$Taiwan Consortium of Emergent Crystalline Materials (TCECM), Ministry of Science and Technology, Taipei 10622, Taiwan}
\affiliation{$^{7}$Research Center for Applied Sciences, Academia Sinica, Taipei 11529, Taiwan}

\begin{abstract}

Scanning tunneling microscopy visualizations of quasiparticle interference (QPI) enable powerful insights into the \textit{k}-space properties of superconducting, topological, Rashba and other exotic electronic phases, but their reliance on impurities acting as scattering centers is rarely scrutinized. Here we investigate QPI at the vacuum-cleaved (001) surface of the Dirac semimetal ZrSiS. We find that interference patterns around impurities located on the Zr and S lattice sites appear very different, and can be ascribed to selective scattering of different sub-sets of the predominantly Zr 4d-derived band structure, namely the \textit{m} = 0 and \textit{m} = $\pm$1 components. We show that the selectivity of scattering channels requires an explanation beyond the different bands' orbital characteristics and their respective charge density distributions over Zr and S lattices sites. Importantly, this result shows that the usual assumption of generic scattering centers allowing observations of quasiparticle interference to shed light indiscriminately and isotropically upon the \textit{q}-space of scattering events does not hold, and that the scope and interpretation of QPI observations can therefore be be strongly contingent on the material defect chemistry. This finding promises to spur new investigations into the quasiparticle scattering process itself, to inform future interpretations of quasiparticle interference observations, and ultimately to aid the understanding and engineering of quantum electronic transport properties.

\end{abstract}

\maketitle

\newpage

\section{I. Introduction}

Observations of quasiparticle interference (QPI) using scanning tunneling microscopy (STM) have become a powerful and increasingly common tool for the characterization of two-dimensional carriers at crystalline surfaces. They have enabled some of the most spectacular discoveries in the field of high temperature superconductivity \cite{Hoffman2002,Hanaguri2009,Aynajian2012} and topological materials \cite{Roushan2009,Zhang2009,Alpichshev2010}. The discovery of topological insulators (TIs) has spurred the prediction and realization of numerous new quantum electronic phases in rapid succession, and exotic quasiparticles in TI, topological crystalline insulator (TCI) \cite{Zeljkovic2015}, Dirac/Weyl semimetal (DSM/WSM) \cite{Jeon2014,Zheng2016,Inoue2016,Batabyal2016}, and nodal-line semimetal materials \cite{Guan2016} have all been subject to investigation using STM imaging of QPI.

Most QPI observations rely on scattering of electronic quasiparticles from scattering potentials usually provided by point defects, or occasionally by atomic steps or magnetic flux vortices induced in superconductors under high magnetic field \cite{Hanaguri2009,Hanaguri2010}. The scattering momentum-transfer vectors connecting initial and final quasiparticle states, \textit{\textbf{q}} = \textit{\textbf{k}}$_{f}$ - \textit{\textbf{k}}$_{i}$, are visualized using Fourier transform imaging of QPI modulations. The relative strength or suppression of various \textit{q}-vectors is often used to infer attributes of the bands which are subject to selection rules governing scattering from initial to final states. Such selectivity has been exploited to explore the spin texture in spin-momentum-locked bands structures of topological and Rashba systems \cite{Roushan2009,ElKareh2013,Steinbrecher2013,Kohsaka2015,Butler2016,Kohsaka2017}, and the symmetry and momentum-dependent sign of the order parameter in unconventional superconductors \cite{Hoffman2002,Hanaguri2009,Hanaguri2010}. Aside from the special case of scattering from magnetic flux vortices, which are known to have scattering properties qualitatively different from simple Coulomb potentials \cite{Hanaguri2010}, impurities have until recently been treated as generic scattering centers, albeit with a relative scattering strength depending on the orbital character of the scattered band, or the different degrees to which impurity lattice sites coincide with a particular band's spatial charge density \cite{Kohsaka2015,Butler2016}. On this view, the presence of any impurities is in principle enough to shed light impartially upon all the allowed scattering processes in \textit{q}-space. 

In this work, we visualize QPI phenomena in the bands at the surface of the Dirac semimetal ZrSiS. Moreover, we take this as a platform from which to examine unusual selectivity in the scattering of quasiparticle bands depending on the type of point defects providing the scattering center. Impurities at Zr and S lattice sites in ZrSiS do not just scatter with different strengths, they selectively scatter different bands. We find that this selectivity is not adequately explained in terms of the orbital characters of bands, and that a more detailed or more exotic explanation is called for. Generally, such impurity dependent scattering phenomena give a hint that the insights allowed by QPI observations might be highly contingent on the particular types of impurities or other point defects allowed by a given material's defect chemistry, and not in a way immediately obvious from the orbital characteristics of the bands in question. Furthermore, the discovery of this phenomenon suggests an avenue towards detailed control of quantum transport \textit{via} control of quasiparticle scattering and lifetimes on the level of specific bands, through choice of substitutional impurities. These issues should motivate further investigation and the development a deeper understanding of the scattering process generally.

The current material in which this effect is observed, ZrSiS, is a Dirac nodal-line semimetal with a number of unusual electronic properties. While TI, TCI, DSM and WSM materials host isolated point-like band crossings of (ideally) linearly dispersive bands, nodal-line semimetals feature crossings between two bands which intersect, without a gap, along a line or a closed loop in the Brillouin zone (BZ) \cite{Schoop2016,Chan2016}. In ZrSiS, a closed diamond shaped nodal-loop-like feature traverses the BZ at energies near the Fermi level, but is in fact gapped by the spin-orbit interaction, while a separate open nodal-line is pinned to the BZ boundary, where it is protected by non-symmorphic symmetry \cite{Schoop2016,Neupane2016}. The bands emanating from the diamond shaped nodal-loop in ZrSiS have linear dispersion over a large energy range, extending over several eV. (The calculated bulk band structure can be found in the Supplementary Information \cite{Supplement}.) Several unusual transport phenomena associated with these band properties have been reported, including very large, anisotropic and non-saturating magnetoresistance \cite{Lv2016,Ali2016,Wang2016,Singha2017,Zhang2018}. For this work, the flatness and stability of the cleaved surface, the relative simplicity of the band structure's constant energy contours (CECs), and the variety and sparsity of points defects, make ZrSiS a neat platform for QPI measurements in general, and scrutinizing the role of defect chemistry in particular.

\section{II. Results}

\subsection{A. Overview of STM and QPI results}

ZrSiS has a quasi-two-dimensional layered structure, depicted in Fig. 1(a), belonging to the space group P4/nmm. Each layer is composed of a Si square net sandwiched by Zr, and further sandwiched by S layers. These quintuple-layer units stack along the \textit{c}-axis and are bonded to each other by the van der Waals interaction. As indicated in Fig. 1(a), cleavage occurs at the van der Waals bonded planes between adjacent S layers, resulting in a square (001) surface lattice of S atoms \cite{Schoop2016,Sankar2016}. A view of the surface-projected positions of each atomic sub-lattice is shown in Fig. 1(b). Single crystals of ZrSiS were grown using a two-step chemical vapor transport method described previously \cite{Sankar2016}, resulting in rectangular platelets as shown in Fig. 1(c).

\begin{figure}[h]
\centering
\includegraphics[scale=1]{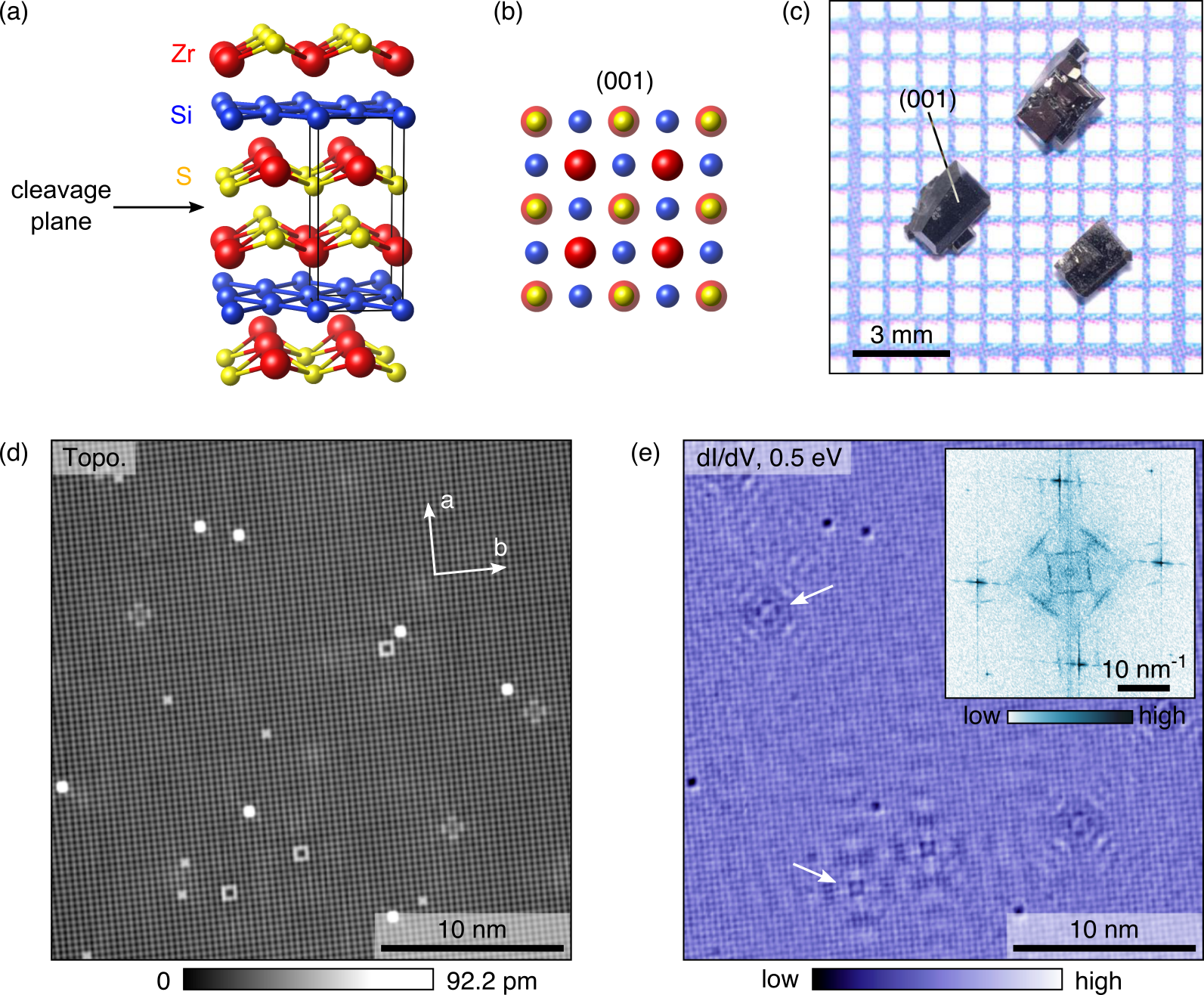}
\caption{\label{fig:1} \textbf{Overview of STM and QPI observations at the cleaved ZrSiS(001) surface.} (a) Crystal structure of ZrSiS, with a cleavage plane between adjacent S square nets indicated. Also shown, in (b), is the top-down view of the (001) surface, and the relative alignment between the Zr, Si and S atomic sub-lattices. (c) Macroscopic view of ZrSiS single crystals, showing large, flat and reflective (001) primary facets. (d) Typical STM topography (taken at \textit{V} = 0.2 V, \textit{I} = 0.3 nA), showing a square surface lattice with various types of impurities. (e) A conductance map at \textit{V} = 0.5 eV in the same field of view, showing highly directional QPI modulations around impurity sites. Impurities of different types clearly cause QPI modulations of different directionality and wavevector. The inset shows the 2D-FFT image with Bragg peaks and QPI signals clearly discernible.}
\end{figure}

For STM measurements, a ZrSiS platelet was cleaved at room temperature under a pressure of $\sim1 \times 10^{-10}$ mbar before transfer into an Omicron LT-STM held at 4.5 K and under a pressure lower than $5 \times 10 ^{-11}$ mbar. STM measurements were performed using a chemically etched tungsten tip, and all d\textit{I}/d\textit{V} data were obtained using a standard lock-in technique with a modulation \textit{V}$_{rms}$ = 10 mV. STM topography taken at the cleaved (001) surface, shown in Fig. 1(d), reveals sparse point defects of different types, some appearing as a small bright `box', and others as a `hash' pattern. Close inspection shows that they are centered at different lattice sites with respect to the observed surface corrugation. Which atomic lattice (Zr, Si, or S) the observed corrugation corresponds to is a necessary reference from which to locate the various point defects, and will be returned to below. Fig. 1(e) shows a d\textit{I}/d\textit{V} image taken at 0.5 eV in the same field of view as (d). This energy has no particular significance other than that it allows easy simultaneous visualization of the QPI signals of interest in the following discussion. At this energy it is found that set-point effects \cite{Li1997,Kohsaka2007,Ziegler2009} have a negligible impact on QPI imaging (see Supplementary Information \cite{Supplement}). Clear, highly directional and coherent QPI modulations are seen near several impurities. Importantly, the difference in orientation of the modulations near impurities of different types indicates that entirely different scattering vectors \textit{\textbf{q}} are involved. This observation, and its implications for QPI measurements in this and other material systems, motivate the remainder of this work.

\subsection{B. Identification of impurity lattice sites}

Our first aim is to identify the nature of the two major types of impurity which cause the markedly different QPI modulations, marked by white arrows in Fig. 1(e). (A more comprehensive categorization and discussion of the observed defects can be found in the Supplementary Information.) Figure 2 details STM, STS and calculated local density of states (LDOS) results which allow us to determine whether the atomic corrugation observed in STM topography corresponds to the S surface lattice, or to the underlying Zr or Si atoms located under the `hollow' or `bridge' sites of the S lattice. This in turn allows the location of each of the various impurities responsible for particular QPI signals. Figure 2(a) shows a typical d\textit{I}/d\textit{V} spectroscopy curve taken at the surface, which is reasonably well reproduced by a plot of the surface LDOS, shown in (b), calculated for a slab model in the framework of density functional theory (DFT). All band structure and charge density calculations presented in this work were performed using the projector-augmented wave (PAW) method \cite{Blochl1994,Kresse1999} as implemented in the VASP package \cite{Kresse1993,Kresse1996,Kresse1996a}, with the LDA exchange-correlation functional \cite{Perdew1981}, and without accounting for spin-orbit coupling (taken to be negligible in ZrSiS). In all cases, a $\Gamma$-centered Monkhorst-Pack \textit{k}-point mesh \cite{Pack1977} was used, and structural optimization was performed with a kinetic energy cutoff of 400 eV and an energy convergence criterion of 10$^{-5}$ eV. 8$\times$8$\times$4, 8$\times$8$\times$1 and 2$\times$2$\times$1 \textit{k}-point meshes were used for the bulk case (see Supplemental Information), the pristine slab, and the slabs hosting impurities (see below), respectively. For calculations of the bands and charge densities near the pristine surface, a 5 quintuple-layer slab model was used. In all slab models the vacuum layer was 24 \AA\enspace thick. The resulting element-resolved contributions to the overall LDOS indicate that it is strongly dominated by Zr states (specifically three of the five Zr 4d states), while the contributions from the S surface lattice and the underlying Si lattice are negligible. (The apparent contraction of observed features along the energy scale in the calculated LDOS is discussed in the Supplementary Information.) These results already hint that tunneling measurements, including STM topography, represent the Zr lattice, and that the S lattice is invisible despite being on top.

\begin{figure}
\centering
\includegraphics[scale=1]{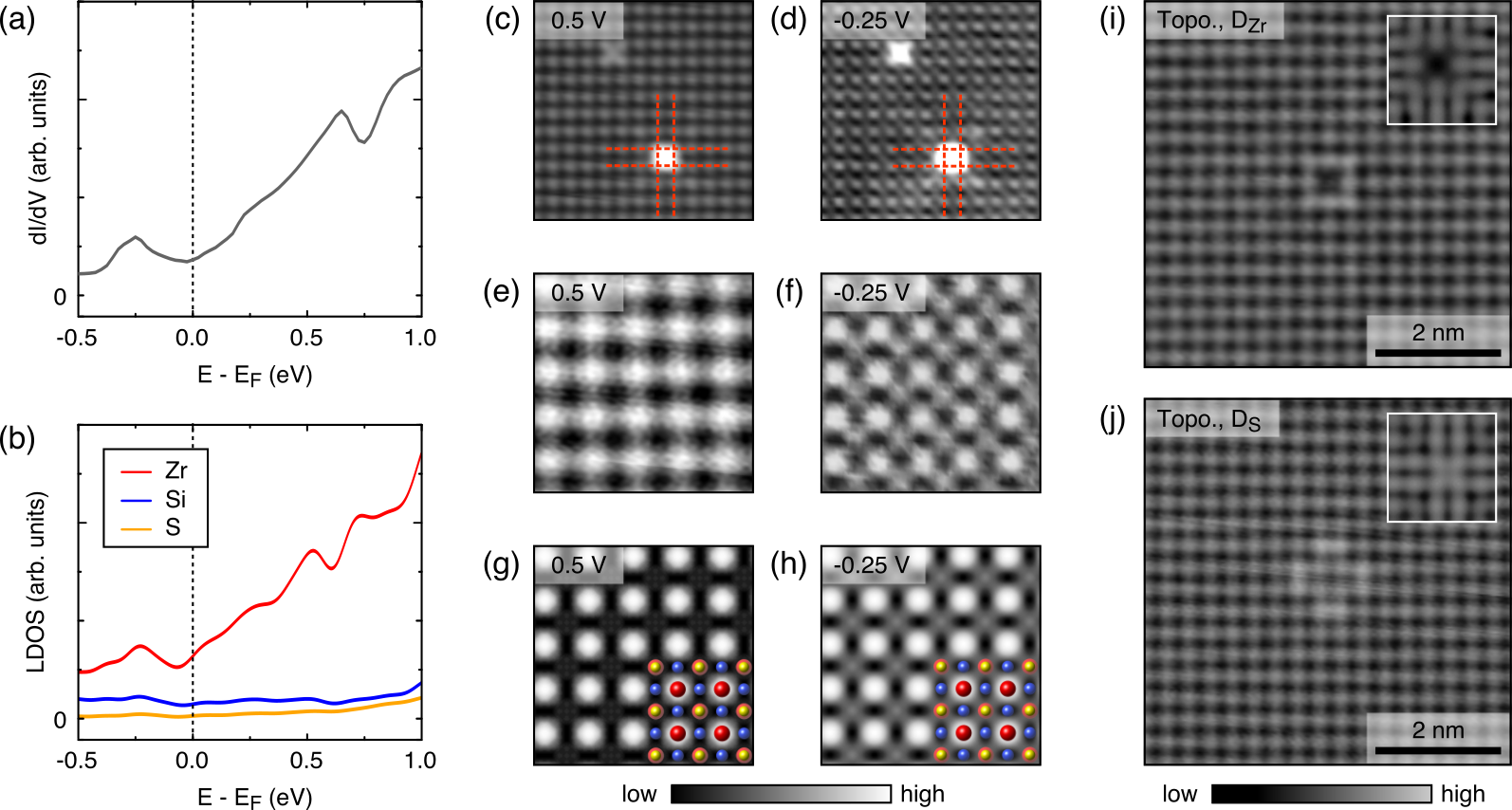}
\caption{\label{fig:2} \textbf{Identification of observed surface lattice and impurity sites.} A typical scanning tunneling spectroscopy curve, shown in (a), corresponds well with the calculated surface projected LDOS, shown in (b). The element-resolved LDOS indicates tunneling measurements, including topography maps, are dominated by Zr-derived states. Voltage-dependent topography maps, taken using \textit{I} = 1 nA, and at \textit{V} = 0.5 eV and \textit{V} = -0.25 eV respectively, are shown in (c) and (d). Taking point defects as a position reference shows that the predominant topographic corrugation belongs to the same atomic sub-lattice in each case. Zoom-in views of a 5$\times$5 unit-cell array at each voltage, (e) and (f), compare well with the corresponding simulated STM images, (g) and (h), in which the atomic positions are known, showing that the most prominent observed corrugation corresponds to the Zr atomic layer. The common types of point defect shown in the topography maps in (i) and (j) can then be identified as D$_{Zr}$ and D$_{S}$ respectively. Simulated STM images of possible candidates for D$_{Zr}$ and D$_{S}$ (modeled as S$_{Zr}$ and Si$_{S}$ substitutions, respectively) are shown in the insets, and reproduce the qualitative features observed in experimental STM maps.}
\end{figure}

Voltage-dependent topography images, acquired with parameters \textit{I} = 1 nA, and at \textit{V} = 0.5 eV and \textit{V} = -0.25 eV, are shown in Fig. 2(c) and (d) respectively. In Fig. 2(d), close inspection reveals that choosing parameters for a very low tip-sample distance allows the imaging of two interleaved lattices simultaneously. They are tentatively identified as the Zr and S lattices, but so far it can not be known which corrugation corresponds to which lattice. Using the defect centers as reference points common to both images, we see that the most prominent corrugation observed at each bias voltage is the same one (as indicated by the red dashed lines). Zoom-in images at each bias, displayed in Figs. 2(e) and (f), provide two cases for comparison with simulated STM maps, in which the positions of each atomic lattice are known from the outset. Simulated STM images were obtained from the partial charge density integrated over the energy range between E$_{F}$ and e\textit{V} (which is noted in each respective image description), using the application of a simple Tersoff-Hamann model described previously \cite{TersoffHamann,Butler2017}. The tip-sample distances are $\sim$5 \AA\enspace and $\sim$2.75 \AA\enspace for the images corresponding to \textit{V} = 0.5 V and \textit{V} = -0.25 V, respectively. These simulated maps, shown in Figs. 2(g) and (h), reproduce the measured topographic corrugations remarkably well, and allow us to attribute the prominent corrugations measured at each bias voltage to the Zr lattice of the crystal. The appearance of the secondary (S) lattice at -0.25 eV is consistent with the fact that, according to the calculated element-resolved LDOS [Fig. 2(b)], the ratio of Zr- to S-derived LDOS, although still greater than unity, is smaller than at 0.5 eV, making the S lattice more amenable to imaging alongside the Zr lattice.

Based on this we ascribe labels to defects centered at the Zr sites (D$_{Zr}$) and the S sites (D$_{S}$), which are exemplified in Figs. 2(i) and (j). Further simulated STM maps, with candidate substitutional impurities included at or near the surface of the slab model, are displayed in the insets of panels (i) and (j) (tip-sample distance $\sim$5 \AA). For the incorporation of impurity atoms, the slab model was repeated in a 5$\times$5 array, and a single substitution was introduced on one side. Substitutions were chosen from among Zr, Si, S and I atoms (I being the transport agent used in crystal growth) and placed at either S or Zr lattice sites. At this point, we do not attribute any measured defect pattern to any specific type of impurity. These comparisons are intended only to verify that simulated STM maps for impurities at particular lattice sites reproduce the qualitative features of the observed D$_{Zr}$ and D$_{S}$. The good qualitative match between experimental and simulated D$_{Zr}$ and D$_{S}$ provide another piece of evidence supporting our identification of the respective impurities' lattice sites.

To provide further confirmation of which sub-lattice is imaged in typical STM topography maps, and thereby also help to confirm the lattice site of each type of impurity, further experimental evidence could be gained by STM imaging of samples with targeted substitutions on the Zr or S lattice sites.

\subsection{C. Interpretation of QPI around individual impurities}

At this point we examine the detailed QPI modulations and associated scattering vectors \textit{\textbf{q}} around each type of impurity individually. In Fig. 3, panel (a) presents the d\textit{I}/d\textit{V} map at an energy of 0.5 eV taken in the vicinity of a D$_{Zr}$ impurity, showing strong and highly coherent QPI modulations. (The simultaneously acquired topography map is shown in the inset). The corresponding 2-D fast Fourier transform (FFT) image in panel (b) shows a scattering channel oriented along the $\overline{\Gamma X}$ direction, which we call \textit{\textbf{q}}$_{1}$, and another (not readily visible in real-space) characterized by the vector \textit{\textbf{q}}$_{2}$ near the Bragg peak. Panels (c) and (d) show the corresponding pair of images for the local QPI around D$_{S}$, showing a similarly strong and highly coherent modulation oriented along the $\overline{\Gamma M}$ direction. It is clear that superimposing both QPI patterns in a single image would effectively reproduce the 2-D FFT shown for the survey of the surface in the inset of Fig. 1(e), because large-scale QPI observations including an ensemble of many different defects capture the aggregate of their induced scattering channels' various contributions in one FFT image.

\begin{figure}
\centering
\includegraphics[scale=1]{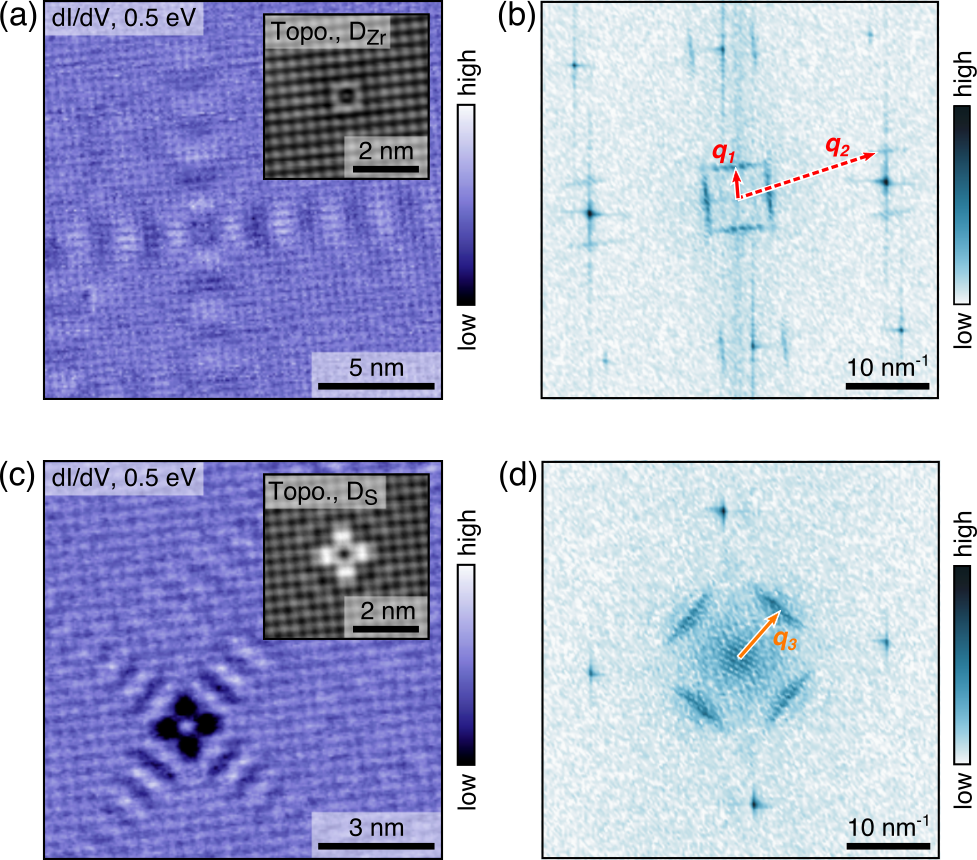}
\caption{\label{fig:3} \textbf{Local QPI observations around D$_{Zr}$ and D$_{S}$ impurities.} (a) A conductance map (\textit{V} = 0.5 V, \textit{I} = 1 nA) showing QPI modulations stemming from a defect centered at a Zr lattice site, with the topography image of the defect pattern shown in the inset (\textit{V} = 0.2 V, \textit{I} = 0.3 nA). (b) The corresponding 2D-FFT image. (c) and (d) The corresponding conductance and 2D-FFT images for a defect centered at a S lattice site, taken using the same parameters as in (a) and (b). Clearly different sets of scattering wavevectors are seen, labelled \textit{\textbf{q}}$_{1}$ and \textit{\textbf{q}}$_{2}$ for D$_{Zr}$, and \textit{\textbf{q}}$_{3}$ for D$_{S}$.}
\end{figure}

We now ascribe scattering vectors \textit{\textbf{q}} to each of the QPI signals visualized in the FFT images of Figs. 1 and 3, with reference to the surface projected band structure and its CEC at the energy of E - E$_{F}$ = 0.5 eV. This CEC is plotted over the range of the 1$^{st}$ surface BZ as in Fig. 4(a), with red and orange arrows marking the transitions satisfying the main scattering vectors \textit{\textbf{q}}$_{1,2,3}$ identified so far. To obtain such CECs, the spectral weight derived from the uppermost three atomic layers of the slab model was sampled using a 320$\times$320 \textit{k}-point mesh. The contribution of the uppermost three layers was chosen to yield results comparable with the highly surface sensitive measurements obtained using STM/STS. Further support for the attribution of each \textit{q}-vector to a transition within each set of bands comes from visualizing the dispersive behavior of each QPI signal. Figs. 4(b) and (c) show the surface electronic band structures plotted along the axes of \textit{\textbf{q}}$_{1}$ and \textit{\textbf{q}}$_{3}$, i.e. along the $\overline{M X M}$ and $\overline{M \Gamma M}$ cuts of the surface BZ, respectively.

Plots of the d\textit{I}/d\textit{V}(\textit{\textbf{q}},E) dispersion were obtained from spectroscopic imaging measurements (with a total measurement time of around 9 hours for each data set) in the same fields of view as Figs. 3(a) and (c). Cuts were taken along the $\overline{\Gamma X}$ axis (parallel to \textit{\textbf{q}}$_{1}$) through the set of FFTs for D$_{Zr}$, and along the $\overline{\Gamma M}$ axis (parallel to \textit{\textbf{q}}$_{3}$) for D$_{S}$, and are presented in Figs. 4(d) and (e). A non-dispersive background signal [an image collected from the minima of d\textit{I}/d\textit{V}(\textit{V})$\rvert _{\textit{q}_{x},\textit{q}_{y}}$ for each pixel] has been subtracted equally from the data at each energy. These cuts display the expected electron-like (and hole-like) dispersive behavior for \textit{\textbf{q}}$_{1}$ (\textit{\textbf{q}}$_{3}$). More quantitative representations are shown in Figs. 4(f) and (g), in which Lorentzian curve fitting is used to extract the peak d\textit{I}/d\textit{V}(\textit{\textbf{q}}) at each energy (open circles and diamonds), for comparison against the peak of the calculated nesting function \textit{N}(\textit{\textbf{q}},E) (indicated by the solid lines), which was obtained by the simple self-convolution of the CEC at each energy. The overall features of the QPI signals \textit{\textbf{q}}$_{1,2,3}$, including the selectivity which is of interest here, are consistent over a large energy range.

\begin{figure}[h]
\centering
\includegraphics[scale=1]{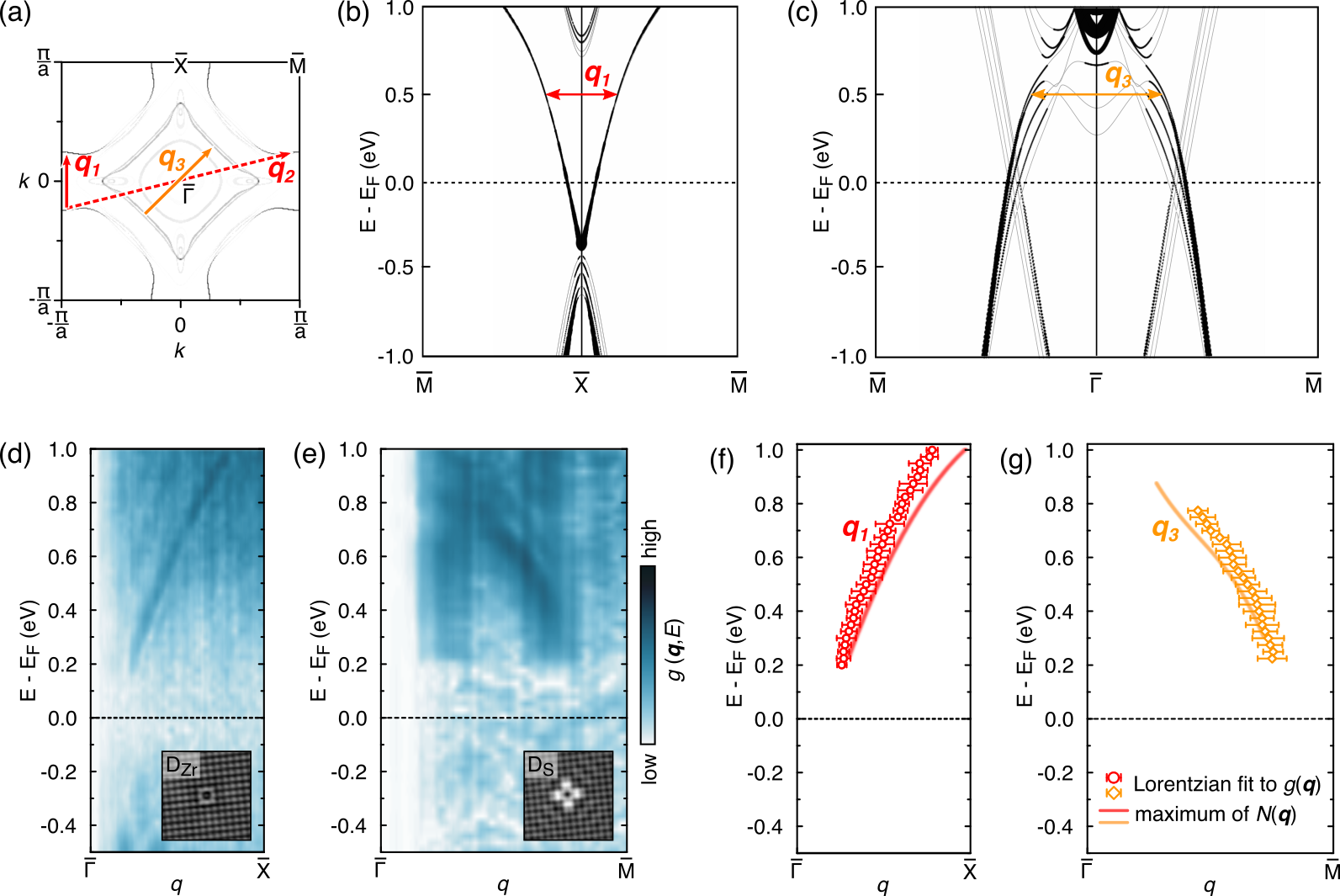}
\caption{\label{fig:4} \textbf{Dispersive behavior of QPI phenomena.} (a) The \textit{q}-vectors described in Fig. 3 are identified within the CEC of the surface-projected bands. The axes of interest are $\overline{M X M}$ and $\overline{M \Gamma M}$. The surface band diagrams (derived from the top three atomic layers) along these axes are shown in (b) and (c) respectively, indicating electron-like dispersion for \textit{\textbf{q}}$_{1}$ and hole-like dispersion for \textit{\textbf{q}}$_{3}$. Experimental QPI dispersion plots along the $\overline{ \Gamma X}$ and $\overline{ \Gamma M}$ cuts are shown in (d) and (e). A comparison of the experimental QPI peaks determined using Lorentzian fitting of d\textit{I}/d\textit{V}(\textit{\textbf{q}}) = \textit{g}(\textit{\textbf{q}}) shown in (d) and (e), with the expected dispersion behavior obtained from the peak along $\overline{ \Gamma X}$ and $\overline{ \Gamma M}$ of \textit{N}(\textit{\textbf{q}}) at each energy, are shown in (f) and (g). The good agreement between measurement and calculation allow us to attribute the \textit{q}-vectors of interest to scattering within particular bands.}
\end{figure}

Having identified the particular bands within which the \textit{q}-vectors are nested, we are interested in the attributes of those bands which distinguish them from each other, and possibly render them susceptible in different degrees to scattering by specific defects. (We postpone for now any discussion of the distinguishing attributes of the defects themselves which allow them to couple more or less strongly to particular bands.) In principle, if the CEC is made up of bands formed from the orbitals of two different constituent elements of a material, then impurities on a particular element's lattice sites may selectively scatter the corresponding orbital sub-set of the CEC. This concept alone does not satisfactorily explain the selective scattering seen here. As is shown in Fig. 2(b), the LDOS and CECs at the ZrSiS surface are formed almost entirely from Zr orbitals. We therefore look to the more detailed orbital make-up of the CEC by decomposing it into its five Zr 4d components.

Figures 5(a--c) show that the total CEC can be decomposed neatly into two sub-sets, the d$_{z^{2}}$ (magnetic quantum number \textit{m} = 0) and d$_{xz}$ + d$_{yz}$ (\textit{m} = $\pm$1) components which form, respectively, a hole-like pocket around the $\overline{M}$ point (previously identified as a surface state \cite{Schoop2016,Neupane2016,Sankar2016}) and an electron-like pocket around the $\overline{\Gamma}$ point (a bulk band stemming from the nodal-loop feature). The d$_{x^{2}-y^{2}}$ component is found to be nearly evenly distributed over the CEC, but significantly weaker than the aforementioned components in the regions which give rise to \textit{\textbf{q}}$_{1,2,3}$, and the d$_{xy}$ component is found to be negligibly small (see Supplemental Information). The corresponding nesting functions \textit{N}(\textit{\textbf{q}}) are shown in (d--f). In (d), \textit{N}(\textit{\textbf{q}}) for the total CEC is computed with the assumption that \textit{m} is conserved in scattering processes. A more detailed discussion of this assumption can be found in the Supplemental Information. Panel (g) shows again the measured QPI patterns for the large field of view shown in Fig. 1, giving the aggregate QPI signal of scattering processes for an ensemble of defects of various types, and panels (h) and (i) show the QPI patterns taken around D$_{Zr}$ and D$_{S}$. A comparison between the predicted and measured QPI patterns for D$_{Zr}$ and D$_{S}$, apparently indicates that they selectively scatter the \textit{m} = 0 and \textit{m} = $\pm1$ sub-sets, respectively.

\begin{figure}
\centering
\includegraphics[scale=1]{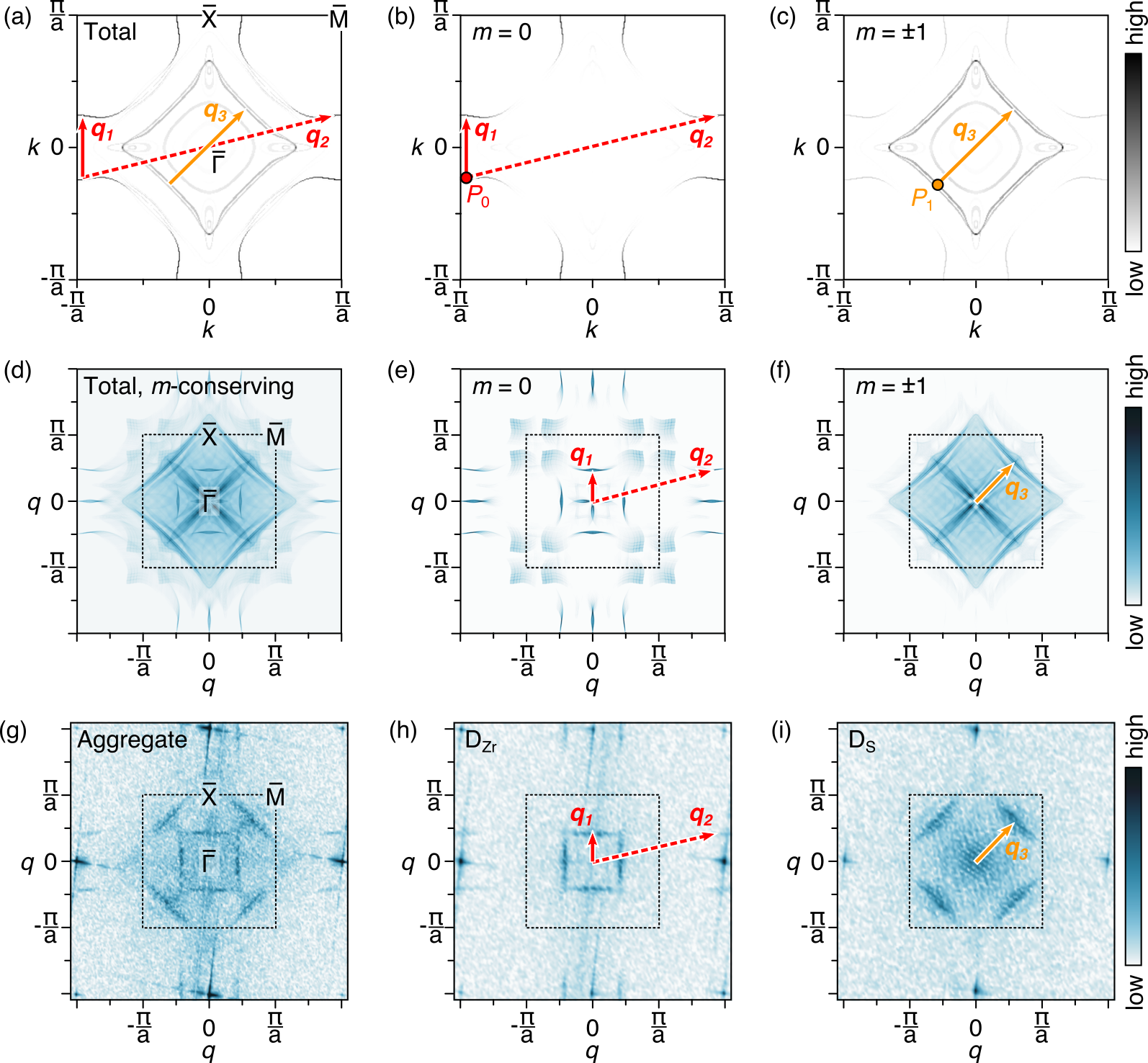}
\caption{\label{fig:5} \textbf{Relation of defect-site-dependent QPI selectivity to the orbital character of bands.} (a) Total CEC, and orbital-resolved CECs for (b) \textit{m} = 0 and (c) \textit{m} = $\pm1$, at E - E$_{F}$ = 0.5 eV, with the prominent scattering wavevectors \textit{\textbf{q}}$_{1,2,3}$ marked. (d-f) The corresponding \textit{N}(\textit{\textbf{q}}) for the total (\textit{m}-conserving) and orbital-resolved CECs. (g-i) Experimental QPI patterns acquired over the large area shown in Fig. 1, and locally around D$_{Zr}$ and D$_{S}$ respectively, as shown in Fig. 3. The QPI measured over a large area gives the aggregate scattering effects of all impurities, which is likely a linear combination of the scattering effects of D$_{Zr}$ and D$_{S}$.}
\end{figure}

\subsection{D. Spatially projected partial charge densities for \textit{m} = 0 and \textit{m} = $\pm$1 dominated bands}

So far, it is still unclear why different sub-sets of the band structure should be susceptible to scattering only by particular impurities. Considering that all the impurities seen in ZrSiS are almost certainly non-magnetic, and are only distinguished by their different positions within the ZrSiS unit cell, the explanation may lie in the real-space charge density distributions of the \textit{m} = 0 and \textit{m} = $\pm$1 dominated components of the band structure. The selectivity might then be explained in terms of the degree to which impurity atoms coincide spatially with regions of high charge density for each band. For example, if the charge density of the \textit{m} = 0 band resides around the Zr lattice sites, and the \textit{m} = $\pm$1 resides elsewhere, then D$_{Zr}$ would disrupt only the \textit{m} = 0 band, causing scattering.

\begin{figure}[h]
\centering
\includegraphics[scale=1]{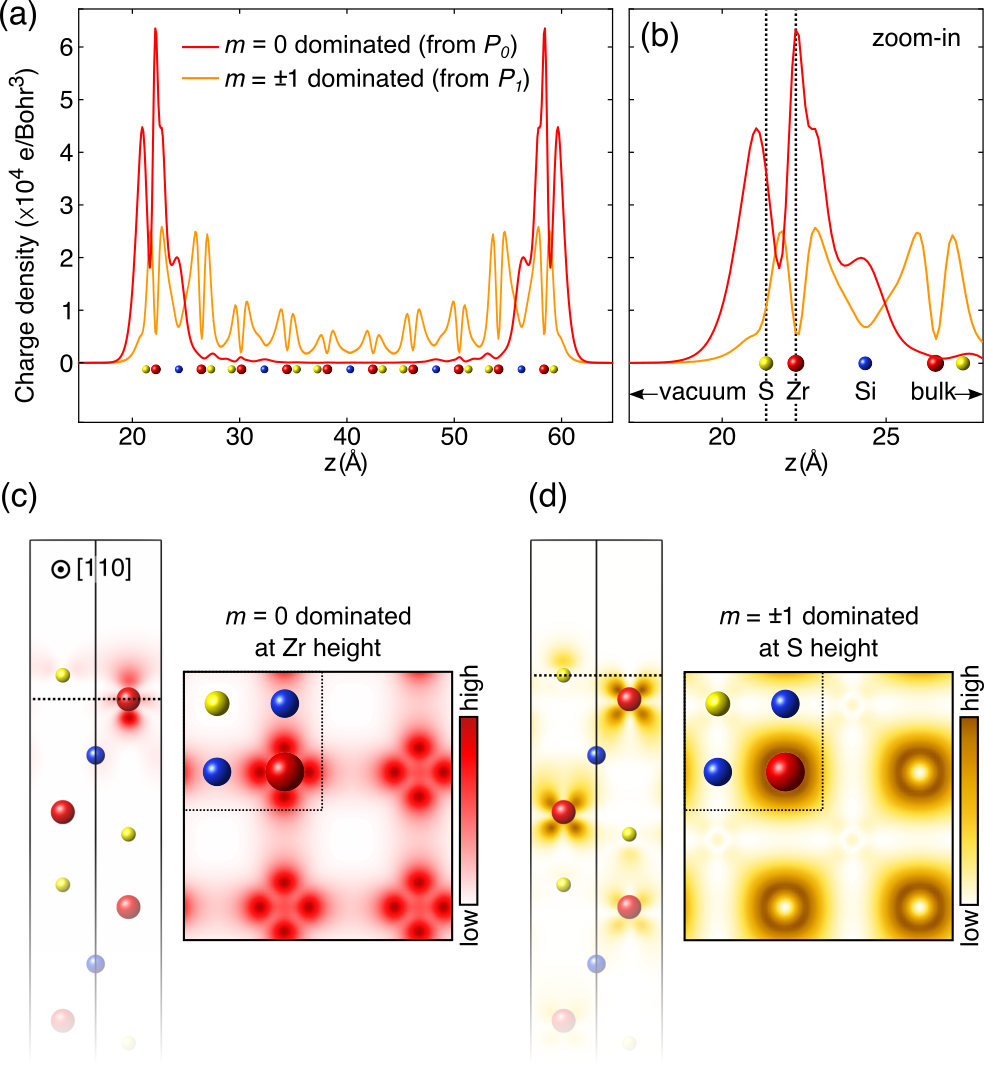}
\caption{\label{fig:6} \textbf{Spatial distribution of partial charge densities for \textit{m} = 0 and \textit{m} = $\pm1$ dominated \textit{k}-points [\textit{P}$_{0}$ and \textit{P}$_{1}$ in Fig. 5].} (a) Calculated charge densities as a function of depth for the slab model. The charge density sampled from the \textit{m} = 0 dominated \textit{k}-point \textit{P$_{0}$} shows the clear characteristics of a surface state, vanishing within the interior of the slab. (b) Zoom-in charge densities in the vicinity of the surface, showing a lack of any particular dominance by either partial charge component around the \textit{z} positions of either the S or Zr lattice sites. (c) and (d) Planar projections of each partial charge component at the \textit{z} height of the S (Zr) lattice site for the \textit{m} = 0 (\textit{m} = $\pm1$) dominated component. Both components appear to have higher intensities around the Zr site.}
\end{figure}

Figure 6 displays various representations of the partial charge distributions, projected back to real-space, of relevant samplings of the \textit{m} = 0 and \textit{m} = $\pm$1 dominated sub-sets of the band structure within an energy range between 0.45 and 0.55 eV. These charge densities were acquired by sampling at the representative points \textit{P}$_{0}$ and \textit{P}$_{1}$ marked in Fig. 5, panels (b) and (c), respectively. The representative components were then projected on the slab super-cell in real-space. Figure 6(a) shows the respective charge densities averaged over the planes perpendicular to the \textit{c}-axis, and plotted as a function of depth throughout a 5 quintuple-layer slab model. The \textit{m} = 0 dominated component appears to be confined to the surface as expected, while the \textit{m} = $\pm$1 dominated component is relatively evenly distributed over the thickness of the slab. A zoom-in view of the surface layer, Fig. 6(b), shows more clearly the respective charge density distributions at the \textit{z} coordinates of the Zr and S layers. There is no particularly strong asymmetry between the charge density near the Zr and S lattices for either \textit{m} = 0 or \textit{m} = $\pm$1. Figures 6(c) and (d) shows the charge densities plotted on the (110)- and (001)-oriented slices through the Zr and S positions. [Since D$_{Zr}$ (D$_{S}$) is the impurity of interest for \textit{m} = 0 (\textit{m} = $\pm$1), the (001) slice is taken at the \textit{z}-coordinate of the Zr (S) layer.] In each viewing orientation both \textit{m} = 0 and \textit{m} = $\pm$1 dominated components are mostly distributed around the Zr site. Hence this does not help to explain how D$_{S}$ scatters either orbital sub-set of the CEC. Nor does it explain why D$_{Zr}$ primarily scatters the \textit{m} = 0 dominated component.

\section{III. Discussion}

The scattered bands appear to exhibit no immediately obvious features which connect either one specifically to the Zr of S lattice sites, either in terms of their elemental orbital character (Zr 4d for all bands), or in terms of their spatial distributions across the lattice. An explanation more focused on the defects' properties, rather than those of the bands may be fruitful instead (in which case band properties such as \textit{m} might only be incidental to the actual mechanism). The form and symmetry of a given defect's Coulomb potential V$_{def}$(\textit{\textbf{r}}) influences the corresponding potential matrix V$_{def}$(\textit{\textbf{q}}) used in the standard T-matrix approach to computing quasiparticle scattering patterns \cite{Zhou2009,Lee2009,Kohsaka2017}. Although the scattering vectors \textit{\textbf{q}}$_{1}$ and \textit{\textbf{q}}$_{3}$ are distinguished by a relative angle of about 45$^{\circ}$, we should not expect V$_{def}$(\textit{\textbf{q}}) for either type of defect to be so sharply anisotropic in a metal with effective screening. Also, from Figs. 3 and 4 it is clear that the potentials V$_{D_{S}}$(\textit{\textbf{q}}) and V$_{D_{Zr}}$(\textit{\textbf{q}}) significantly overlap in terms of the magnitudes of the resulting vectors \textit{\textbf{q}}$_{1,2,3}$. These observations indicate that sharply differing defect-dependent forms of V$_{def}$(\textit{\textbf{q}}) alone are unlikely to explain the selectivity. This situation seems to call for a more detailed, comprehensive or perhaps more exotic explanation beyond simple spatial charge densities. For example, since quasiparticle scattering involves a momentum transfer \textit{\textbf{q}} which may entail electron-phonon interactions, looking at how particular modes within the phonon band structure relate to displacements of the Zr and S sub-lattices might be one avenue to connect different \textit{q}-vectors to different lattice sites.

At this stage, without a comprehensive framework to understand and predict band-selective scattering by different impurities, it is not clear whether this phenomenon is general, confined to some sub-set of material systems, or (most unlikely) confined solely to ZrSiS. If it is more general, this could have importance for QPI measurements in emergent materials in the future. Many QPI observations have exploited mechanisms of selectivity governing scattering channels, such as the spin texture of bands, or the variations in sign of the superconducting order parameter, variables grounded in \textit{k}-space. These observations rely on scattering from various defects providing a predictably uniform view on \textit{q}-space. The results described here raise the possibility that in particular cases, such observations may in fact be highly contingent on the particular types of defect present (a materials chemistry problem). At worst, one can imagine a hypothetical case where the defect chemistry leads to only a sub-set of the band structure being accessible to QPI observations, potentially limiting conclusions which can be drawn about the \textit{k}-space band properties. In future experiments relying on QPI, the effect described here may need to be taken into account while interpreting QPI results to examine other forms of selectivity for scattering channels, especially in materials with very sparse defects, and in fact may be exploited when designing QPI experiments in the future. More broadly, this finding hints at an avenue towards the detailed engineering of quasiparticle scattering, lifetimes and transport properties on the level of specific targeted bands, by choice of substitutional impurities. Ultimately, it is hoped that this finding will help to spur developments of a more detailed theoretical understanding of the quasiparticle scattering process itself.

\vspace{5mm}

During the final stages of preparation of this manuscript, we became aware of another work based on QPI observations at the ZrSiS surface. See Ref. \cite{Lodge2017}.

\section{Acknowledgements}
We thank C.-C. Su, T.-M. Chuang, G.-Y. Guo, Y. Kohsaka, and T. Hanaguri for invaluable discussions. This work was supported in part by the Ministry of Science and Technology of Taiwan through the following grants: MOST 102-2119-M-002-004, and MOST 105-2119-M-002-013.

\end{document}


\title{Supplementary Information\protect\\Quasiparticle Interference in ZrSiS: Strongly Band-Selective Scattering Depending on Impurity Lattice Site} 

\author{Christopher J. Butler$^{1,}$\footnote{These authors contributed equally to this work.}$^{,}$\footnote{cjbutler@ntu.edu.tw}}
\author{Yu-Mi Wu$^{1,}$\footnotemark[1]}
\author{Cheng-Rong Hsing$^{2}$}
\author{Yi Tseng$^{1}$}
\author{Raman Sankar$^{3,4}$}
\author{Ching-Ming Wei$^{2,}$\footnote{cmw@phys.sinica.edu.tw}}
\author{Fang-Cheng Chou$^{4,5,6}$}
\author{Minn-Tsong Lin$^{1,2,7,}$\footnote{mtlin@phys.ntu.edu.tw}}

\affiliation{$^{1}$Department of Physics, National Taiwan University, Taipei 10617, Taiwan}
\affiliation{$^{2}$Institute of Atomic and Molecular Sciences, Academia Sinica, Taipei 10617, Taiwan}
\affiliation{$^{3}$Institute of Physics, Academia Sinica, Taipei 11529, Taiwan}
\affiliation{$^{4}$Center for Condensed Matter Sciences, National Taiwan University, Taipei 10617, Taiwan}
\affiliation{$^{5}$National Synchrotron Radiation Research Center, Hsinchu 30076, Taiwan}
\affiliation{$^{6}$Taiwan Consortium of Emergent Crystalline Materials (TCECM), Ministry of Science and Technology, Taipei 10622, Taiwan}
\affiliation{$^{7}$Research Center for Applied Sciences, Academia Sinica, Taipei 11529, Taiwan}

\maketitle
\thispagestyle{empty}

\begin{figure}
\centering
\includegraphics[scale=1]{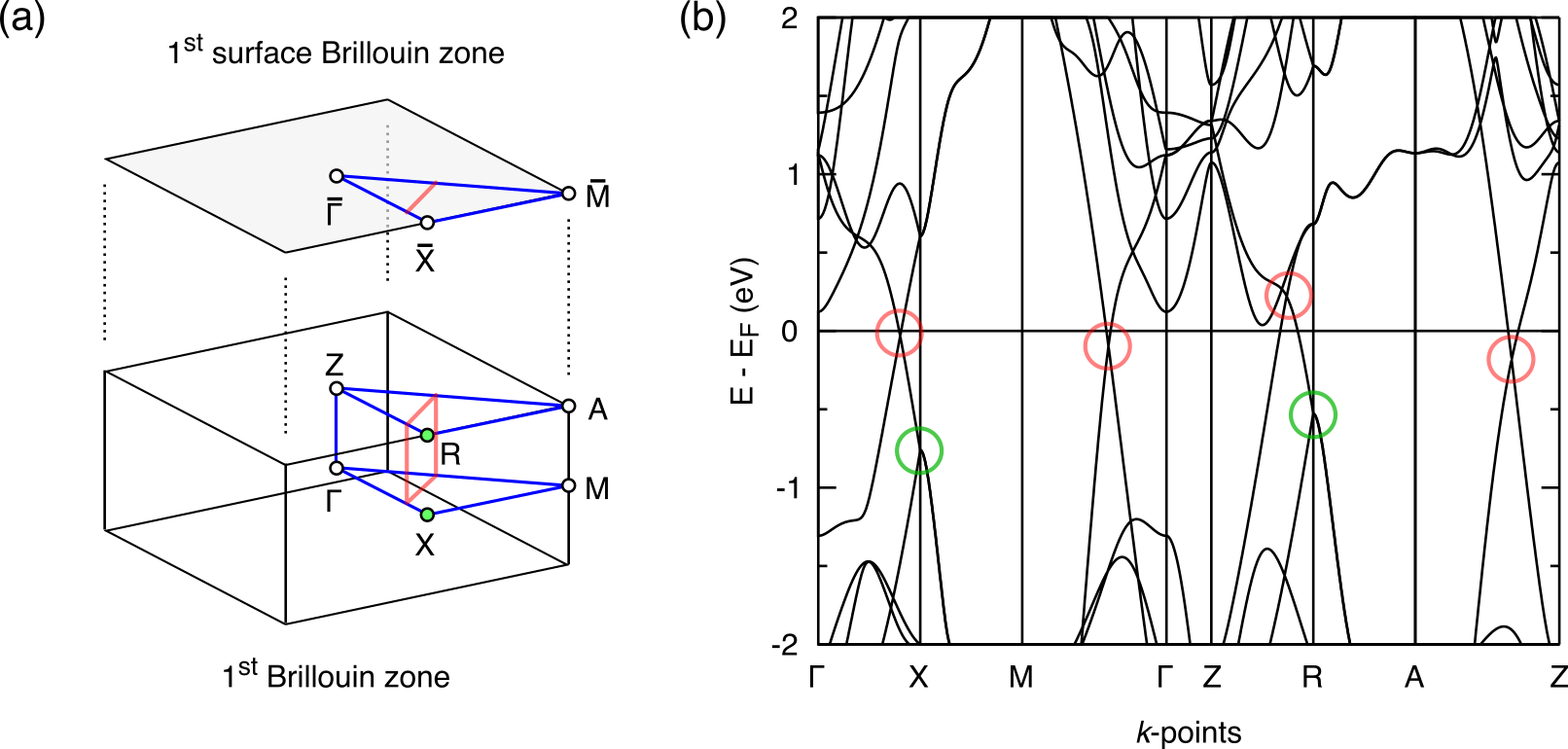} 
\caption{\label{fig:1} (a) The bulk and surface projected Brillouin zones (BZs), also showing the position of the bulk Dirac nodal-loop and its surface projection (pink lines). The paths along high-symmetry axes, on which the bulk and surface band structures are sampled, are shown as blue lines. (b) The bulk band structure, calculated as described in the main article. The points where the high-symmetry axes intersect the bulk Dirac nodal-loop are circled in pink, and the Dirac nodal-line protected by non-symmorphic symmetry (see Ref. [17]) is circled in green.}
\end{figure}

\pagebreak

\begin{figure}
\centering
\includegraphics[scale=1]{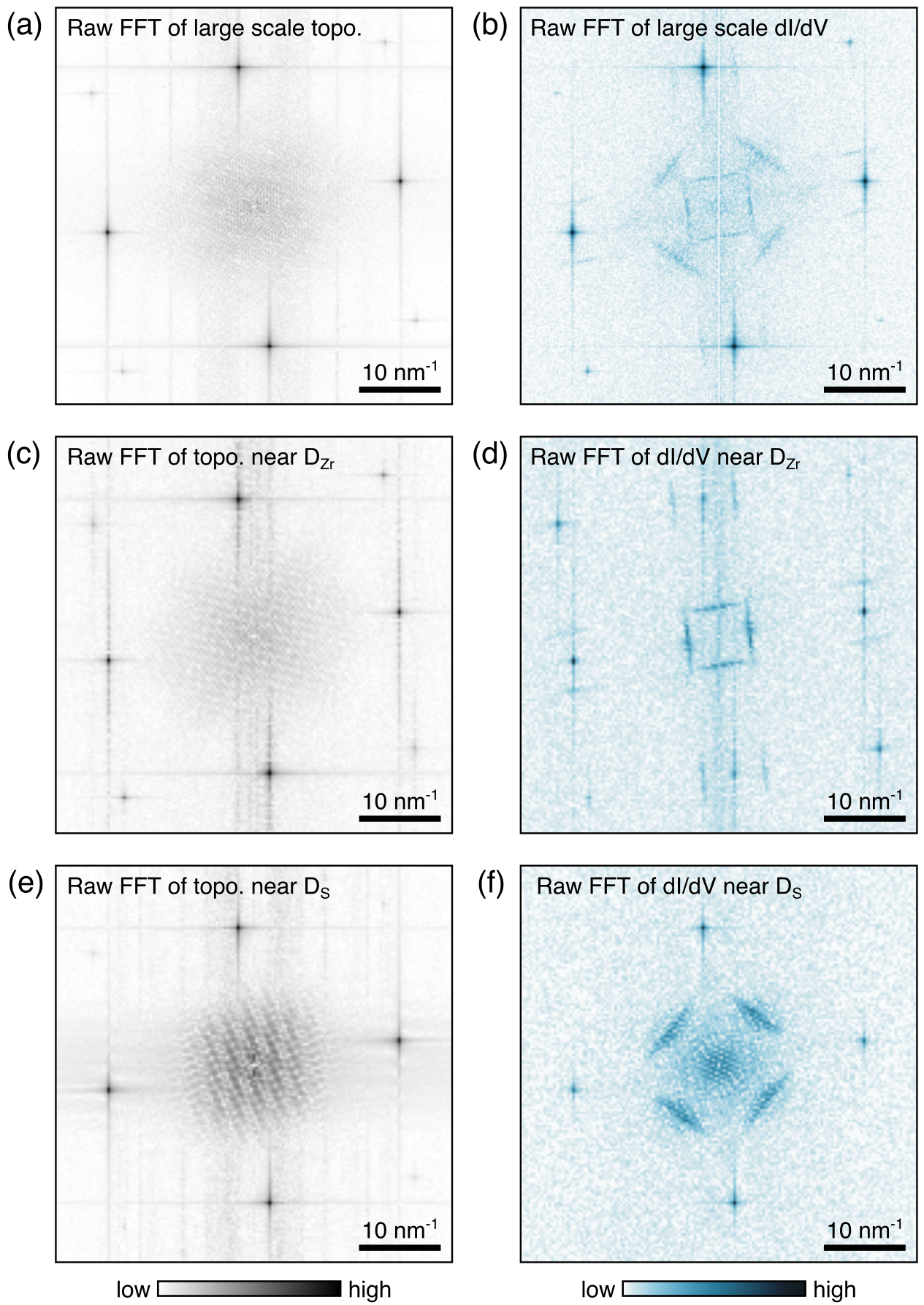}
\caption{\label{fig:2} Raw Fourier transforms of simultaneously acquired topography and d\textit{I}/d\textit{V} images. Artificial QPI-like components of the topography images which might be imprinted onto the d\textit{I}/d\textit{V} images due to the `set-point effect' (see Refs. [29--31])  are seen to be negligible.}
\end{figure}

\pagebreak

\begin{figure}
\centering
\includegraphics[scale=1]{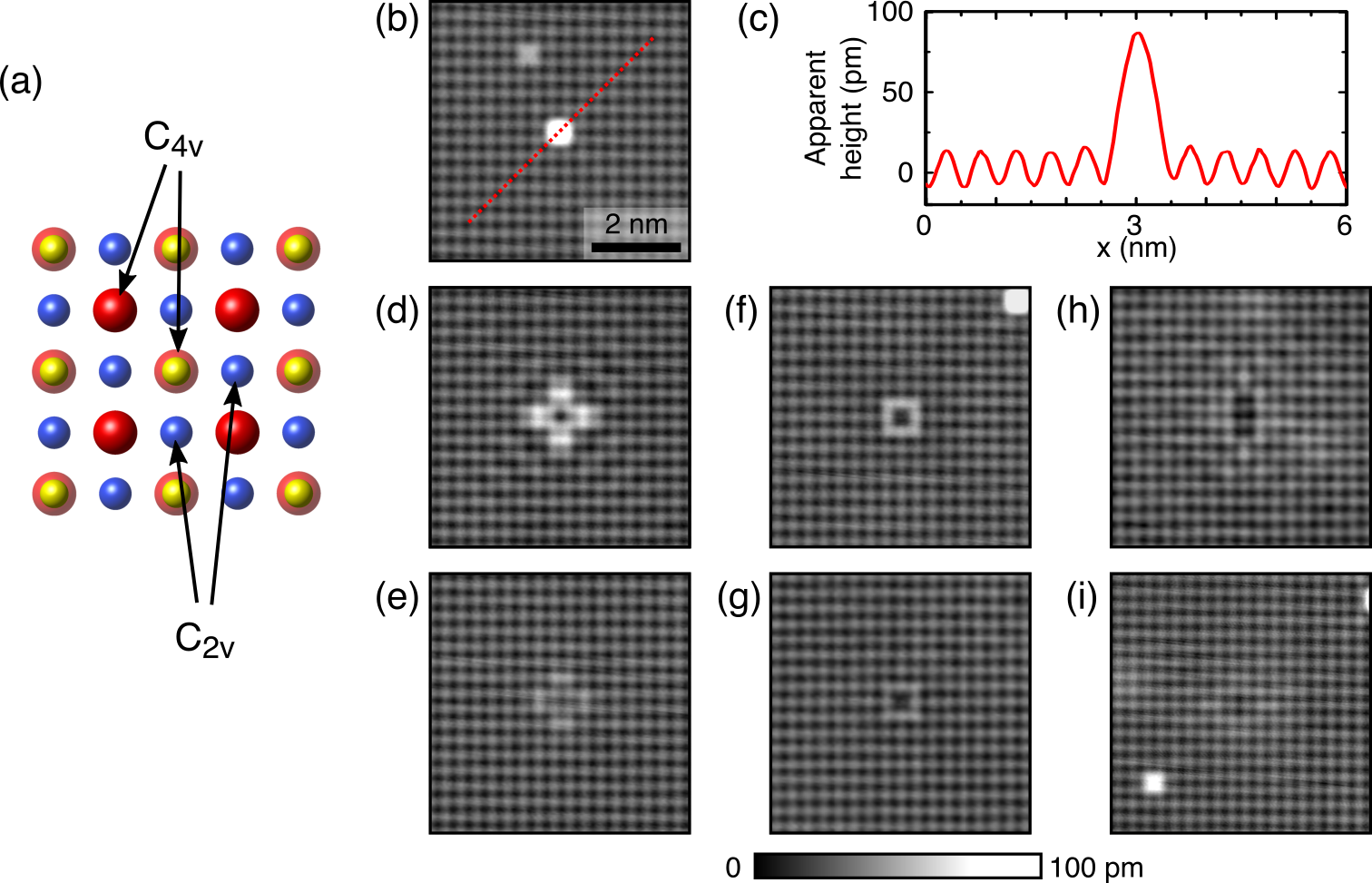}
\caption{\label{fig:3} (a) A top-down view of the (001) surface of ZrSiS. The two-dimensional point groups associated with the Zr, Si and S lattice sites are marked. (b) A prominent protrusion previously identified as an adatom sitting atop a hollow site of the surface lattice. As the observed corrugation actually corresponds to Zr, this protrusion is located at or on top of a S site, and so is unlikely to be an adatom. It is more likely to be a substitution at a S site, such as Zr$_{S}$, with the large apparent height of $\sim$1 \AA\enspace [shown in the line profile in (c)] due to being embedded in the otherwise undetected S surface lattice. (d--i) Two varieties each of D$_{S}$, D$_{Zr}$ and D$_{Si}$. Defects having C$_{2v}$ symmetry are most likely located at the Si sites (although interstitial impurities with this symmetry may be possible). For defects at each lattice site, instances with varying topographic height are found. The upper panels [(d), (f) and (h)] show the most prominent examples found, while the lower panels [(e), (g) and (i)] show weaker examples. These are interpreted as impurities of different atomic species at each given site, causing differing degrees of charge transfer between the impurity and the host lattice. All STM images here were acquired using \textit{V} = 0.2 V and \textit{I} = 0.3 nA.}
\end{figure}

\pagebreak

\begin{figure}
\centering
\includegraphics[scale=1]{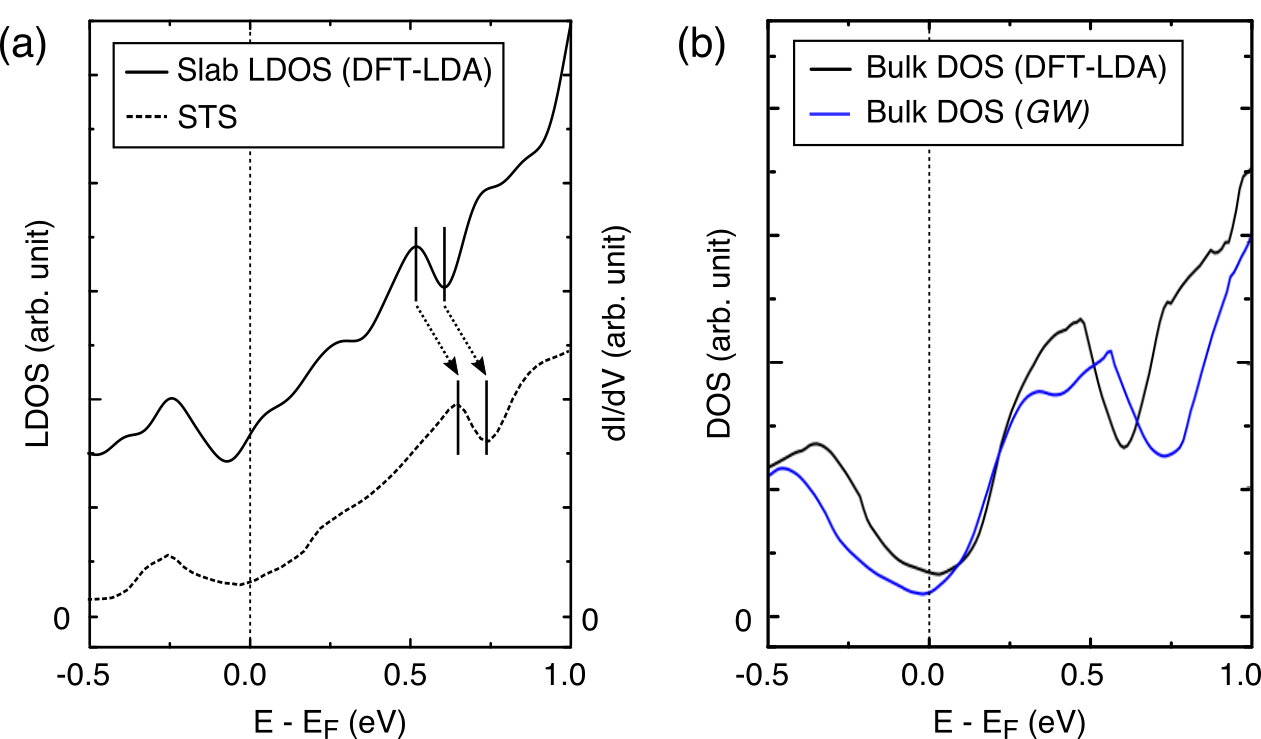} 
\caption{\label{fig:4} (a) Measured STS versus calculated LDOS (using DFT) for the ZrSiS surface, as shown in Figs. 2(a) and 2(b). The main discrepancy is a $\sim$150 meV downward energy shift in the LDOS as compared to the measurement. (b) The bulk DOS calculated using DFT versus that calculated using the \textit{GW} approach with a 12$\times$12$\times$6 \textit{k}-point mesh. The \textit{GW} result shows features consistent with the DFT result, but with an upward energy shift roughly proportional to E - E$_{F}$, and comparable with the energy shift shown in (a) for energies around 0.6 eV. These results indicate that DFT results underestimate the energy scale of features near E$_{F}$ for both bulk DOS and surface LDOS. This also helps to explain the small mismatch between measured and predicted QPI dispersion shown in Figs. 4(f) and 4(g).}
\end{figure}

\begin{figure}
\centering
\includegraphics[scale=1]{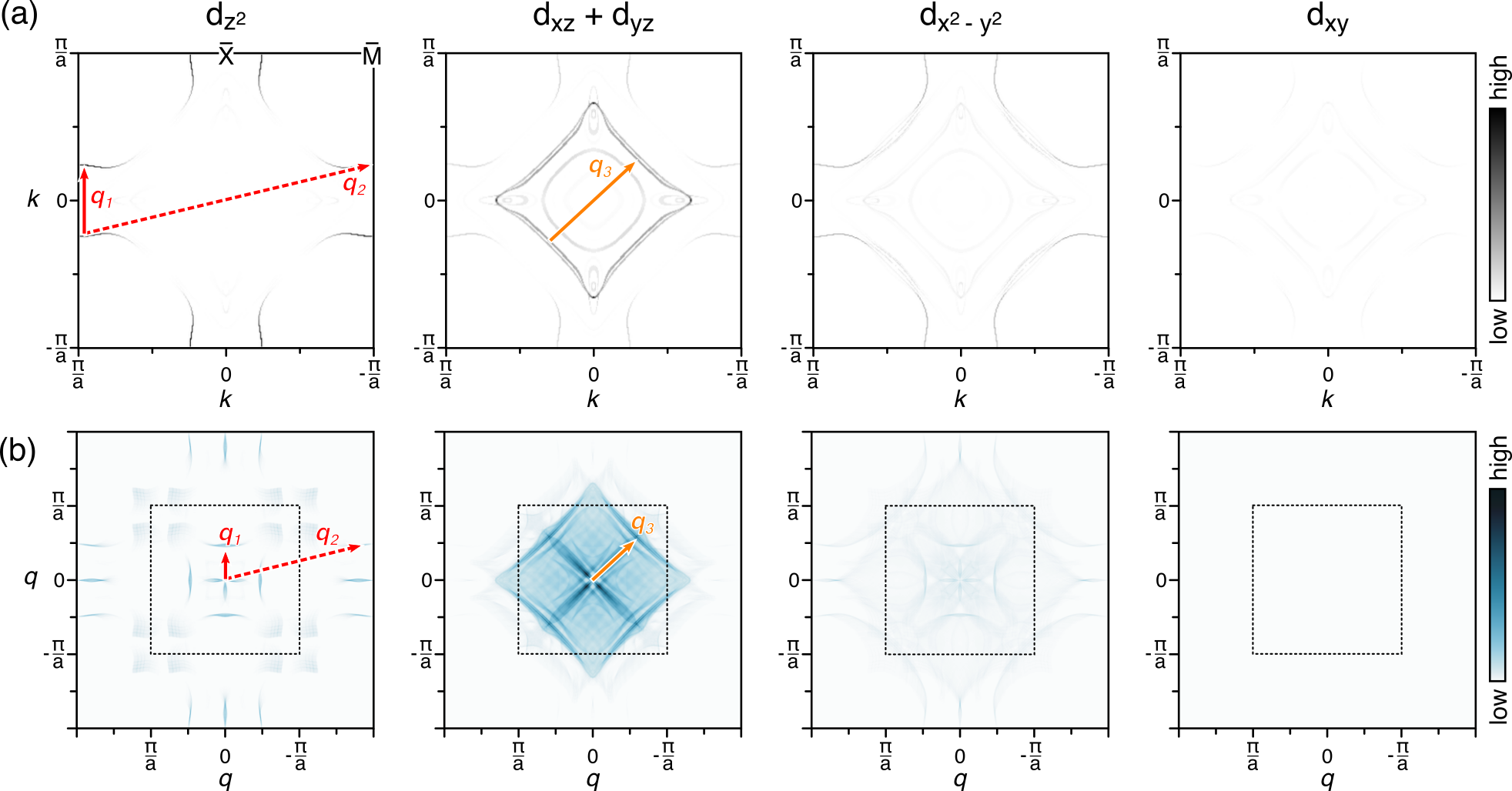}
\caption{\label{fig:5} (a) Orbitally decomposed CECs at 0.5 eV, with spectral weights normalized with respect to each other, and (b) the corresponding nesting functions. Relevant scattering vectors \textit{\textbf{q}$_{1,2,3}$} are marked. The QPI signals expected for the d$_{x^{2}-y^{2}}$ and d$_{xy}$ components are negligible in comparison to those for d$_{z^{2}}$, d$_{xz}$ and d$_{yz}$.}
\end{figure}

\clearpage

\begin{figure}
\centering
\includegraphics[scale=1]{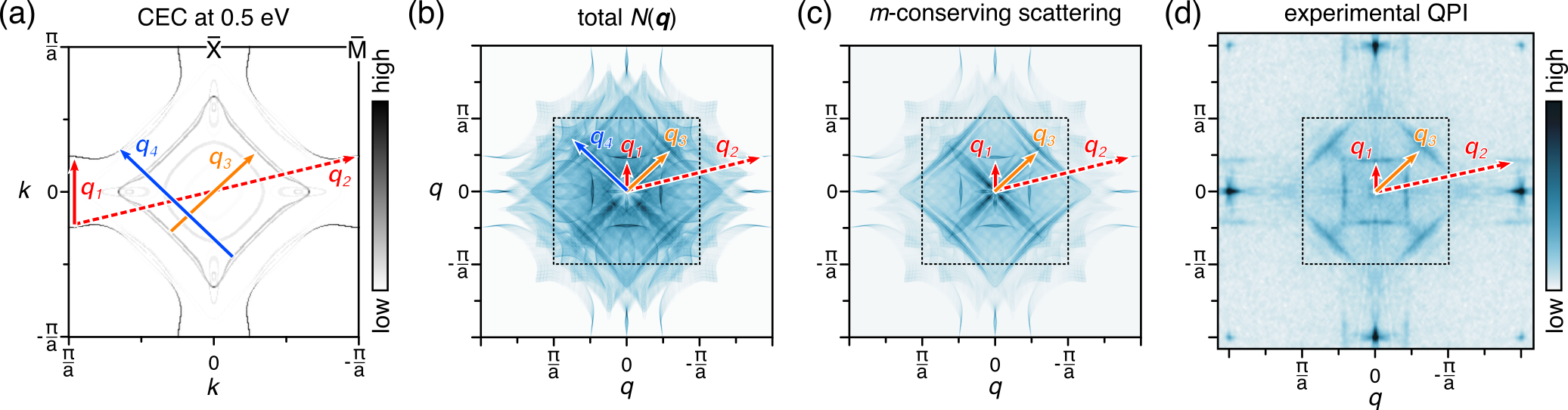}
\caption{\label{fig:6} (a) The total CEC at 0.5 eV, plotted over the surface BZ. The scattering vectors \textit{\textbf{q}$_{1,2,3}$} discussed in the main article are shown, along with a scattering vector \textit{\textbf{q}$_{4}$}, which connects the central electron pocket (\textit{m} = $\pm$1) to the corner hole pocket (\textit{m} = 0). (b) The total nesting function \textit{N}(\textit{\textbf{q}}) for the CEC, with all scattering vectors marked. (c) The \textit{m}-conserving nesting function, obtained by summing the nesting functions computed separately for each orbital component. (d) The symmetrized FFT image of QPI shown in Fig. 1(e) of the main article. We interpret the improved prediction of the measured data by the image in (c) as evidence of suppression of \textit{\textbf{q}$_{4}$} and similar scattering events, and an overall conservation of \textit{m}.}
\end{figure}

\section{Supplementary discussion}
\textbf{Conservation of quantum number \textit{m} in quasiparticle scattering.} Figure S4 above shows (a) the CEC at 0.5 eV, and comparisons of calculated nesting functions which (b) do not, and (c) do conserve \textit{m}, to the experimental data (d).  The experimental QPI is apparently consistent with the suppression of \textit{m}-changing scattering events. This may be expected (at least for non-magnetic defects which cannot impart angular-momentum to the final quasiparticle state) according to the same rationale as for suppression of spin-flip scattering which has been demonstrated in QPI measurements of spin-polarized bands in topological and Rashba systems (see Refs. [4--6, 14--16]). However, here we stress that the general conservation on \textit{m} remains an assumption, and is not strictly demonstrated by the results presented in this work: As each type of defect selectively scatters a certain sub-set of the CEC, and the underlying mechanism for selectivity is not yet known, the orbital character of the bands might be only incidental to the selectivity (and this might be true in general, not only in the current system). In future, and in continually emerging systems suitable for QPI observations, it is still worthwhile to remain attentive to the possibility of observing \textit{m}-changing scattering events.